\RequirePackage{ifpdf}
\ifpdf 
\documentclass[pdftex]{sigma}
\else
\documentclass{sigma}
\fi

\usepackage[all]{xy}
\begin{document}

\allowdisplaybreaks

\renewcommand{\PaperNumber}{047}

\renewcommand{\thefootnote}{$\star$}

\FirstPageHeading

\ShortArticleName{Some Remarks on the KP System of the
Camassa--Holm Hierarchy}

\ArticleName{Some Remarks on the KP System \\ of the Camassa--Holm
Hierarchy\footnote{This paper is a contribution to the Proceedings
of the Workshop on Geometric Aspects of Integ\-rable Systems
 (July 17--19, 2006, University of Coimbra, Portugal).
The full collection is available at
\href{http://www.emis.de/journals/SIGMA/Coimbra2006.html}{http://www.emis.de/journals/SIGMA/Coimbra2006.html}}}

\Author{Giovanni ORTENZI~$^{\dag^1\dag^2}$}

\AuthorNameForHeading{G. Ortenzi}

\Address{$^{\dag^1}$~Dipartimento di Matematica
Politecnico di Torino,\\
$\phantom{^{\dag^2}}$~Corso Duca degli Abruzzi 24, 10129 Torino,
Italy}

\Address{$^{\dag^2}$~Dipartimento di Matematica e Applicazioni
Universit\`a di Milano Bicocca,\\
$\phantom{^{\dag^1}}$~Via R. Cozzi 53, 20125 Milano, Italy}

\EmailDD{\href{mailto:giovanni.ortenzi@unimib.it}{giovanni.ortenzi@unimib.it}}

\ArticleDates{Received October 31, 2006, in f\/inal form January
22, 2007; Published online March 13, 2007}

\Abstract{We study a Kadomtsev--Petviashvili system for the local
Camassa--Holm hierarchy obtaining a candidate to the
Baker--Akhiezer function for its f\/irst reduction generalizing the local Camassa--Holm. We focus our
attention on the dif\/ferences with the standard KdV-KP case.}


\Classification{37K10; 35Q53}

\section{Introduction}
The Camassa--Holm (CH) equation with zero critical velocity
\cite{CamHol}
\begin{gather*}
u_t -u_{xxt}=6u_xu-4u_{xx}u_x-2uu_{xxx}
\end{gather*}
and his multi f\/ields extensions
\cite{AvdL,CLZ,Greg,FLP,I,LP,SM1,SM2,S} are widely studied
integrable systems because they show a number of properties
dif\/ferent  from the standard Gelfand--Dikii systems such as
Korteweg--de Vries (KdV). Its study is important to a deep
understanding of inf\/inite-dimensional integrability.

One of the most important dif\/ferences between CH and KdV is
that, even if CH has the same dispersionless part of  the KdV
equation
\begin{gather*}
u_t=-u_{xxx}+6uu_x,
\end{gather*}
it does not admit tau structure in the sense of Dubrovin's
classif\/ication scheme {\cite{DubZh}}. Classically, for the
Kadomtsev--Petviashivili (KP) equation and its reductions, the tau
structure is related to the fact that the Baker--Akhiezer function
$\psi$ of the hierarchy satisf\/ies a bilinear
relation~\cite{DJKM,SS,SW}.

The bi-Hamiltonian method allows the interpretation of the KP
system as a generalization of the conservation laws for  the
corresponding Noether currents. This construction \cite{FMP}
starts from the Riccati equation related to the system. In the KdV
case the equation
\begin{gather}
\label{RiccKdV} h_x+h^2=u+z^2, \qquad \text{with}\qquad
h(z):=z+\sum_{i \geq -1} \frac{h_i}{z^i},
\end{gather}
is the relation satisf\/ied by the conserved densities $h_i$ of
KdV. Using the Noether theorem one can associate to every
conserved density $h_i$ a current $H^{(3)}_i$ such that
$\partial_t h_i = \partial_x H^{(3)}_i$. Every commuting symmetry
in the KdV hierarchy is related to a dif\/ferent current
 and then, for a~symmetry of KdV involving
 the time ${\partial}/{\partial t_s}=\partial_s$, it holds
$\partial_s h_i = \partial_x H^{(s)}_i$. Using the
generator~$h(z)$ we can collect the currents into a set of
generators of currents satisfying
\begin{gather}
\label{Noe}
\partial_s h(z) = \partial_x H^{(s)}(z).
\end{gather}
At this stage this set of equations is simply a way to rewrite the
equations of the KdV hierarchy.

The amazing property of the currents $H^{(s)}$ is that they belong
to a very particular space called $H_+$. This is the space
generated by the linear span on $C^{\infty}(S^1,\mathbb{R})$ of
the Fa\`a di Bruno polynomials $h^{(n)}:=(\partial_x + h)^n \cdot
1$ of the conserved density  $h$. There is a unique way to write
every current $H^{(s)}$ as $H^{(s)}(x,z)=\sum\limits_{i=0}^s
c_i(x) h^{(i)}(x,z)$.

Without account of (\ref{RiccKdV}), the equation (\ref{Noe})
becomes an inf\/inite dif\/ferential system in an inf\/inite
number of f\/ields. This is equivalent to the KP equation
presented by the Japanese school. This system is a linear f\/low
on a suitable Grassmannian whose $H_+$ is the positive part. A key
property~\cite{CFMPBrasil} of $H_+$ for the relation between KP
and Grassmannian f\/lows is that this space is invariant under the
action of the operators $(\partial_s+H^{(s)})$:
\begin{gather}
\label{dentroKP} (\partial_s + H^{(s)}) H_+ \subset H_+.
\end{gather}
Thanks to (\ref{dentroKP}) the KP system can be written as
\begin{gather}
\label{CSlKdV} (\partial_s +
H^{(s)})H^{(r)}=H^{(s+r)}+\sum_{i=1}^{s}H^r_i
H^{(s-i)}+\sum_{i=1}^{r}H^s_i H^{(r-i)}.
\end{gather}
If we forget the Fa\`a di Bruno rule for construction of  the
space $H_+$, the currents $H^{(s)}$ are simply Laurent series in
$z$ and a collection $\{ H^{(s)} \}_{s \geq 0}$ is related
(\cite{FMP}) to a point of the positive part~$H_+$ of the Sato
Grassmannian $\mathcal{W}$ def\/ined  e.g.\ in \cite{SW}. The
equation (\ref{CSlKdV}) describes now a~f\/low on $\mathcal{W}$
which admits as reduction the KP system imposing the relation
between the currents and the Fa\`a di Bruno polynomials. It is
called the \emph{central system} for KP. The time $t_1$ of the
central system can be identif\/ied with the variable $x$ of KP
because the conservation of the linear momentum of KP equation.

This construction is a hint for the existence of  bilinear
relations \cite{FMP} and one wonders whether the same scheme can
be applied to generalizations of CH hierarchy.

The study of KP equation for CH is interesting in this direction.

In \cite{CLOP} the authors show the existence of a KP equation
constructed starting from local symmetries of the CH equation. In
this paper we plan to pursue  these ideas further. After a review
of the results presented in \cite{CLOP}, we study the KP-CH system
obtaining a candidate for the Baker--Akhiezer function of the
integrable reductions of the system. Next we consider the problem of the reduction from KP-CH
to the local Camassa--Holm hierarchy. By means of this process we
obtain a 3-f\/ield integrable system which generalizes the local
symmetries for CH.

\section{The full CH hierarchy}
It was known \cite{CamHol,KM} that CH is bi-Hamiltonian  on the
vector space $C^{\infty}(S^1,\mathbb{R})$ by means of the Poisson
pencil
\begin{gather*}
P_{z^2}=P_0-z^2 P_1=\frac{1}{2}\partial_x-\frac{1}{2}\partial_x^3
-z^2 \left( -m\partial_x-\partial_x m \right), \qquad z \in
\mathbb{R},
\end{gather*}
where $m=(1-\partial_x^2)u$ and we call $z^2$ the pencil parameter
for later convenience. The theory of bi-Hamiltonian systems (see
e.g.~\cite{MCFP}) shows that the conserved quantities for CH are
the coef\/f\/icients  of the potential of an exact 1-form $v$ in
the  kernel of the Poisson pencil. If
$H(z)=\sum\limits_{k=-\infty}^{+\infty}H_k z^{-k}$ satisfy
$P_{z^2} dH(z)=0$ then one can construct the Lenard--Magri
recursion $P_0 dH_k = P_1 dH_{k+2}$. This chain implies that all
$H_k$ are in involution  w.r.t. both the Poisson pencils.

Every element $v(z)$ in $\mathrm{Ker} P_{z^2}$ satisf\/ies the
equation:
\begin{gather}
\label{eqnperv} \frac{1}{4} v^2 -\frac{1}{2}v_{xx}v
+\frac{1}{4}v_x^2+ z^2 m v^2   = f(m,z),
\end{gather}
where $f(m,z)$ satisf\/ies $f_x=0$. It turns out that $v$ is an
exact 1-form if $f$ does not depend on~$m$. Without loss of
generality, we can put $f(z)={z^2}/{4}$.

The equation (\ref{eqnperv}) can be solved iteratively developing
$v$ in $z$. In order to f\/ind the potential~$H(z)$ we evaluate
$v$ on the generic vector $\dot{m}$:
\begin{gather*}
\langle v, \dot{m} \rangle \buildrel(\ref{eqnperv}) \over{=}
\int\limits_{S^1}v  \frac{d}{dt} \biggl(
\frac{z^2}{4v^2}+\frac{1}{2z^2}\frac{v_{xx}}{v}
-\frac{1}{4z^2}\frac{v_{x}^2}{v^2}-\frac{1}{4z^2} \biggr)\,dx
\\ \phantom{\langle v, \dot{m} \rangle}
{}= \frac{d}{dt} \int\limits_{S^1} \frac{1}{2v}\,dx
 + \int\limits_{S^1} \partial_x \biggl( \dot{v_{x}} + \frac{v_x \dot{v}}{v} \biggr)\,dx
= \frac{d}{dt} \int_{S^1} \frac{1}{2v}\,dx.
 \end{gather*}
Therefore   $ H(z)= \int_{S^1}({1}/{2v})\,dx$ is a generator of
conserved quantities.

Let us focus now our attention on the density $h$ of $H$. It is
obviously def\/ined up to a~normalization and up to a total
derivative. Paying attention to these `gauge' choices one can put
\begin{gather*}
h = z \frac{1}{2v} + \partial_x( \ln \sqrt{v} )
\end{gather*}
and then the equation (\ref{eqnperv}) becomes
\begin{gather}
\label{Riccati} h_x+h^2 = \frac{1}{4}+z^2 m.
\end{gather}
By construction, all the solutions of this equation def\/ine
conserved quantities for CH (\cite{CLOP,Len,R} and, in the context
of the inverse spectral problem, \cite{C,CmK}).

The possible solutions of (\ref{Riccati}) are two functions,
depending on the essential singularity point of $h(z)$. The
f\/irst one, $h(z)=\sum\limits_{i=-1}^{+\infty} h_i z^{-i}$, has
its essential singularity in $0$. It corresponds to the
Lenard--Magri recursion starting from the Casimir of $P_1$ and all
the quantities are local in $m$.

The second one, $k(z)=\sum\limits_{i=0}^{+\infty} k_i z^{i}$,
contains CH and it has an essential singularity at inf\/inity. It
corresponds to the Lenard--Magri recursion starting from the
Casimir of $P_0$, and it has an inf\/inite number of  quantities
nonlocal  in $m$.

The  calculation of these solutions is not dif\/f\/icult but a
little bit involved. It is performed explicitly in \cite{CLOP}. We
now simply list and comment the results.
\subsection*{The local hierarchy}
The local hierarchy  is related to the solutions of
(\ref{Riccati}) of the form
$h(z)=h_{-1}z+h_0+{h_{1}}/{z}+{h_{2}}/{z^2}+\cdots$:
\begin{gather*}
h_{-1}=\sqrt{m}, \qquad h_0=(\ln(m^{-{1}/{4}}))_x,
\\
h_1=\frac{1}{8\sqrt{m}}-\frac{1}{8}\frac{m_{xx}}{\sqrt{m^3}}+\frac{5}{32}\frac{m_x^2}{\sqrt{m^5}},
\\
h_2=\biggl(-\frac{1}{16m}+\frac{1}{16}\frac{m_{xx}}{m^2}+\frac{5}{64}\frac{m_x^2}{m^3}\biggr)_x,\quad
\ldots.
\end{gather*}
The conserved quantities $H_{i} =  \int h_{i-1}\,dx$ for $i=0,1,2,
\dots $ are nontrivial only when $h_{i-1}$ are not a total
derivative, i.e. when $i$ is even. Using only the even conserved
quantities we obtain $H(z^2)= \sum\limits_{i \geq 0} H_{2i}
z^{-2i}$.

We can iterate the Lenard--Magri chain def\/ining an inf\/inite
number of vector f\/ields $X_{2i} = P_0 d H_{2i} = P_1 d H_{2i+2}$
pairwise commuting and we identify $X_{2i}=\partial_{2i} m$. The
f\/irst one, already presented in \cite{CamHol}, is the {\it local
Camassa--Holm equation}
\begin{gather}
\label{locCH}
\partial_0 m =(\partial_x-\partial_x^3)\frac{1}{4\sqrt{m}}.
\end{gather}
This part of the hierarchy generates the negative f\/lows of CH.
\subsection*{ The nonlocal hierarchy}
The nonlocal hierarchy is related to the solutions of
(\ref{Riccati}) of the form $k(z)=k_{0}+k_{-1}z + k_{-2}
z^2+\cdots$:
\begin{gather*}
k_0=\frac{1}{2},\\
k_{-1}= 0,  \\
k_{-2}= (1+\partial_x)^{-1}m=u-u_x,\\
k_{-3}=0, \\
k_{-4}=-(1+\partial_x)^{-1}((1+\partial_x)^{-1}m )^2 = -u^2-u_x^2 + \text{total derivatives},  \\
k_{-5}=0, \\
k_{-6}=2(1+\partial_x)^{-1}((1+\partial_x)^{-1}
((1+\partial_x)^{-1}m )^2(1+\partial_x)^{-1}m )
\\ \phantom{k_{-6}}{}
=u^3+uu_x^2+\text{total derivatives},\quad \ldots .
\end{gather*}
We remark that one can continue the iteration thanks to the
invertibility of the operator $1+\partial_x$ in the space of
smooth periodic functions \cite{CLOP}. The densities are
increasingly non-local in $m$, but the f\/irst three are local in
the f\/ield $u = (1-\partial_x^2)^{-1}m$.

As in the local case one can def\/ine conserved quantities
$K_{-i}=\int k_{-i}  \, dx$ for $i=0,1,2, \dots $. Taking only the
nontrivial ones, $i=2,4,6,\dots$, we obtain $K(z^2)=
\sum\limits_{i \geq 1} K_{-2i} z^{2i}$.

We can iterate the Lenard--Magri chain def\/ining an inf\/inite
number of vector f\/ields $Y_{-2i} = P_0 d K_{-2i} = P_1 d
K_{-2i+2}:=\partial_{-2i} m$ pairwise commuting. The f\/irst
nontrivial vector f\/ield is the conservation of the linear
momentum $\partial_{-2} m = P_1 d K_{-2}= P_0 d K_{-4}= -m_x$. The
second one is the standard Camassa--Holm equation with null
critical velocity
\begin{gather*}
\partial_{-4} m =P_1dK_{-4}=-(m\partial_x+\partial_xm)(1-\partial_x^2)^{-1}(-2u+2u_{xx})
=4mu_x+2m_xu =P_0dK_{-6}
\end{gather*}
that is, with $t_{-4}=t$,
\begin{gather*}
u_t-u_{xxt}= 6u_xu-4u_{xx}u_x-2uu_{xxx}.
\end{gather*}
The conserved quantities $H_i$ and $K_{-i}$  commute with each
other by the bi-Hamiltonian construction. Therefore we can collect
the two iteration chain into a unique one def\/ining the full CH
hierarchy
$$
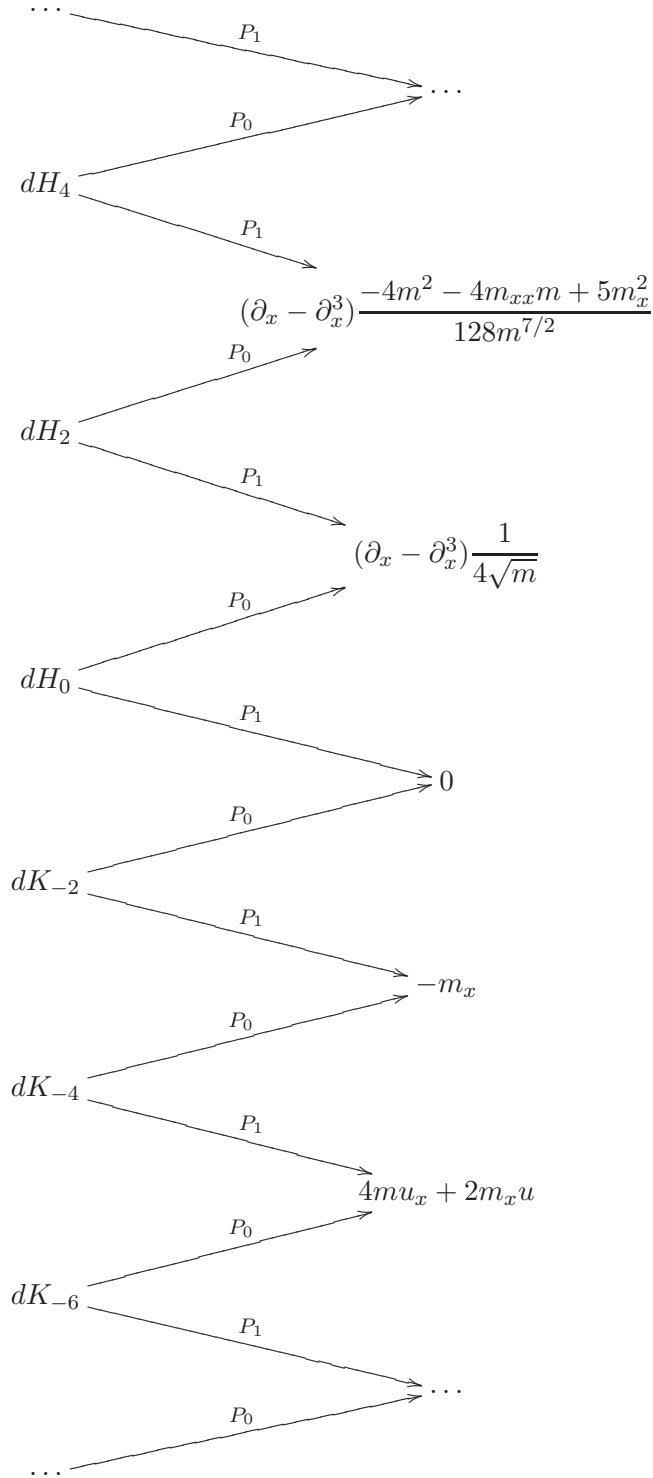
\xymatrix{\dots \ar[rrd]^{P_1}  &&\\
                               && \dots\\
 dH_4\ar[rrd]^{P_1} \ar[rru]^{P_0} &&\\
             && (\partial_x-\partial_x^3)\dfrac{-4m^2-4m_{xx}m+5 m_x^2}{128 m^{7/2}}\\
 dH_2\ar[rrd]^{P_1} \ar[rru]^{P_0} &&\\
             &&(\partial_x-\partial_x^3)\dfrac{1}{4\sqrt{m}}\\
 dH_0\ar[rrd]^{P_1} \ar[rru]^{P_0} &\\
             && 0 \\
dK_{-2}\ar[rrd]^{P_1} \ar[rru]^{P_0} &&\\
             &&-m_x\\
 dK_{-4}\ar[rrd]^{P_1} \ar[rru]^{P_0} &&\\
             &&4mu_x+2m_xu  \\
 dK_{-6}\ar[rrd]^{P_1} \ar[rru]^{P_0} &\\
             && \dots \\
\dots \ar[rru]^{P_0}& }
$$
\vspace{-3mm}

\section{The Noether currents}
 In the KdV case the KP system has a simple interpretation as a
generalization of the Noether relations (\ref{Noe}) between
conserved densities and related currents \cite{FMP}.
 To understand if such a~construction is still possible, we write down  these currents in our case.
In the case of local conserved quantities the calculation is
already performed in \cite{CLOP} and gives the result:
\begin{gather}
\label{Jreg} J^{(s)} = \sum_{i=0}^{s} \biggl(
-\frac{1}{2}\partial_x v_{i}(z^{s-i+2})+v_{i}(z^{s-i+2} h) \biggr)
= \frac{z^{s+3}}{2}+O(z), \qquad s\geq 0.
\end{gather}
With similar computations, using
\begin{gather*}
\partial_{-2s} m = - P_{z^2} \sum_{j=1}^{s} z^{2j-2s-2} dK_{-2j}
\end{gather*}
and $w(z) =z dK(z)$, one can show that, in the nonlocal case, the
currents are:
\begin{gather*}
J^{(-s)} = \sum_{j=1}^{s} \biggl( \frac{1}{2}\partial_x w_{-j}
(z^{j-s-1})- w_{-j} (z^{j-s-1}k) \biggr) =
\frac{z^{-s-2}}{2}+O(1), \qquad s\geq 1.
\end{gather*}
Although it seems to possess the same structure, these two
families of currents are really dif\/ferent. The local currents,
as in the KdV case, are elements of a particular space called
$J_+$. It is the span on the periodic smooth functions of the
Fa\`a di Bruno polynomials
\begin{gather}
\label{FdBz2} h^{(n)}=(\partial_x + h)^n z^2, \qquad n \geq 0.
\end{gather}
It holds
\begin{proposition}
\label{currente} The currents $J^{(2s)}$, with $s\geq 0$, are
elements of $J_+$.
\end{proposition}
\begin{proof} (See \cite{CLOP}.)
Thanks to the representation (\ref{Jreg}), it suf\/f\/ices to show
that $z^{2i} $ and $z^{2i} h $ are elements of $J_+$ for all
$i\geq 1$. First of all, $z^2=h^{(0)}\in J_+$ and $z^2
h=h^{(1)}\in J_+$ by def\/inition of~$J_+$. Moreover, the Riccati
equation (\ref{Riccati}) multiplied by $z^2$,
\begin{gather}
\label{Riccatiz} h^{(2)} = \frac{z^2}{4}+z^4 m,
\end{gather}
shows that $z^4=1/{m} (h^{(2)}-({1}/{4}) h^{(0)})\in H_h$. Acting
with $(\partial_x+h)^n$ on both sides of (\ref{Riccatiz}) we prove
the statement.
\end{proof}
For every $s\geq 0$, the current $J^{(s)}$ has a essential
singularity in zero and a pole of order $s+3$  at inf\/inity. The
space which contains these currents has to be generated by Laurent
series with similar characteristics. Actually, in the proof  of
the previous proposition, a fundamental role is played by the fact
that the conserved density $h(z)$, which is the seed for the Fa\`a
di Bruno polynomials, diverges both at $0$ (essential singularity)
and at $\infty$ (simple pole).

For every $s < 0$, the current $J^{(s)}$ has a essential
singularity at inf\/inity and a pole of order $-s-2$  at zero.
However the related generator of conserved quantities $k(z)$ is
regular at $0$. Therefore  $k(z)$ cannot generate the analogue of
$J_+$ for the nonlocal currents by means of the previous method.

 As we will see in the following, the existence of this space is a key ingredient for the construction  of a
KP-CH system.

\section{The KP-CH system}
Let us then concentrate on the local currents $J^{(s)}$. They can
be characterized in a unique way by the following two properties:
\begin{gather*}
1)\ J^{(s)}=\dfrac{1}{2} z^{s+3}+O(z), \qquad 2)\ J^{(s)} \in J_+.
\end{gather*}
These currents are not a basis of the $J_+$ space. To obtain a
basis one has to add the constant~$z^2$ so that  $J_+=\langle
z^2,J^{(n)} \rangle_{n \geq 0}$. Assuming that $h$ is an arbitrary
Laurent series of the form
\begin{gather*}
h(z)=h_{-1} z+\sum_{i=0}^{+\infty}\frac{h_i}{z^{i}},
\end{gather*}
where the coef\/f\/icients $h_i$ are not constrained by the
Riccati equation, we can def\/ine the currents~$J^{(s)}$, for all
$s\geq 0$, imposing the two above-mentioned properties. We
def\/ine the $s$-th equation of the {\em local KP-CH system\/}
\cite{CLOP} as
\begin{gather}
\label{KP-CH}
\partial_s h=\partial_x J^{(s)},\qquad s\geq 0.
\end{gather}
It is an evolution equation in an inf\/inite number of f\/ields
given by the coef\/f\/icients $h_{-1}$, $h_0$, $h_1$, $\dots $  of
$h$. The main dif\/ference between this system and the standard KP
system {\cite{KMM}} is that the density $h$ is not an element of
the space  $J_+$. Nevertheless the equations (\ref{KP-CH}) can be
written using only elements of  $J_+$
\begin{gather*}
\begin{array}{rl}
\partial_s h^{(1)}   = z^2 \partial_x  J^{(s)}.
\end{array}
\end{gather*}
These equations are nonlinear in the space $J_+$ because they
involve the product of two elements of $J_+$. This nonlinearity in
another dif\/ference between KP-CH and standard KP.

Directly from the equations of motion we easily see that this
system possesses an inf\/inite number of conserved quantities.
Their densities are all the coef\/f\/icients $h_i$ of $h(z)$.

It is well known that the KP equation in $2+1$D can be obtained
simply combining in a~suitable way the f\/irst equation of the KP
system in inf\/inite f\/ields. The analogue of the $2+1$
dimensional KP equation can be obtained also in our case. The
smallest closed dif\/ferential subsystem of evolution equation
involves the f\/irst 5 f\/ields $h_{-1}, \dots, h_3$
\begin{gather*}
\partial_0 h_{-1}=\partial_x \dfrac{h_{1}}{2h_{-1}},
\qquad
\partial_0 h_{0}=\partial_x \dfrac{h_{2}}{2h_{-1}},
\qquad
\partial_0 h_{1}=\partial_x \dfrac{h_{3}}{2h_{-1}},
\\
\partial_1 h_{-1}=\partial_x \biggl( \dfrac{h_{2}}{h_{-1}} + \dfrac{1}{h_{-1}}\partial_x \dfrac{h_{1}}{2h_{-1}}\biggr),
\\
\partial_1 h_{0}=\partial_x \biggl( \dfrac{h_{3}}{h_{-1}} + \dfrac{h_1^2}{2h_{-1}^2}
+\frac{1}{h_{-1}}\partial_x \dfrac{h_{2}}{2h_{-1}}\biggr).
\end{gather*}
After some simple algebraic manipulation one reduces this system
to
\begin{gather*}
2u_t=\biggl( \dfrac{w}{u} \biggr)_x,
\\
u_y=2v_t+\biggl( \dfrac{u_t}{u} \biggr)_x,
\\
v_y=2w_t+2\dfrac{w}{u}u_t+\biggl( \dfrac{v_t}{u} \biggr)_x,
\end{gather*}
where $h_{-1}=u$, $h_0=v$, $h_1=w$ and $t_0=t$, $t_1=y$.

\subsection{The evolution of the currents}
We study now the evolution of the currents $J^{(s)}$. In the
standard KP case they are related to the Sato linear f\/lows on
the Grassmannian \cite{FMP}. For the KP-CH equation the situation
is somewhat dif\/ferent. By def\/inition, the space $J_+$ is
invariant w.r.t. $\partial_x+h$ but the invariance is lost w.r.t.
$\partial_s+J^{(s)}$. Actually the action of $\partial_s+J^{(s)}$
of the generic element of the basis of $J_+$ is
\begin{gather}
\label{esco}
(\partial_s+J^{(s)})h^{(k)}=(\partial_s+J^{(s)})(\partial_x+h)^k
z^2= z^2 (\partial_x+h)^k (\partial_s+J^{(s)}) \subset z^2 J_+.
\end{gather}
\begin{remark}
As we have seen in the introduction, the analogue for KP of the
space $J_+$ is the space $H_+$. The main dif\/ference between
these two spaces is that in the KP-CH case the space~$J_+$ is not
invariant under the action of the operators
$(\partial_s+J^{(s)})$.
\end{remark}

Using arguments similar to those used in \cite{FMP}
it is easy to show, for example in the case
$z^2 J_+ \subset J_+$ studied in the next paragraph,
the commutativity of the flows generated by
$\{ \partial_s \}_{s \geq 0} $. Therefore, in the same case,
we can introduce the function $\psi$ defined by
$\partial_s \ln \psi = J^{(s)}$ and by $\partial_x \ln \psi = h$.
Such a function exist also in the KP case and, in that case \cite{CFMPBrasil},
it is the Baker--Akhiezer function.

We conjecture  that the generalizations of $\psi$ are
the Baker--Akhiezer functions for the integ\-rable reductions of KP-CH.

\section{On the KP-CH reductions}
To recover the local CH hierarchy from (\ref{KP-CH}), one has to
impose on $h$ the constraint given by the Riccati equation
(\ref{Riccati}). Therefore all the f\/ields $h_i$ can be written in
terms of $m$ and its $x$-de\-rivatives. Thus the local KP-CH
system (\ref{KP-CH}) reduces to the local CH hierarchy. As in
the Gelfand--Dikii cases this reduction is a stationary reduction
because the Riccati equation  imply  $J^{(1)}=\frac{z^4}{2}$ and
the triviality of the evolution along $t_1$. From
Proposition~\ref{currente} it also follows that all $t_{2s+1}$ are
stationary. However, contrarily to the KdV case, the constraint given
by Riccati equation (\ref{Riccati}) is not equivalent to
$J^{(1)}=\frac{z^4}{2}$. The local Camassa--Holm equation is a strange
reduction of a system obtained in its turn as a reduction from the
KP-CH. Under the  constraint
\begin{gather}
\label{constr} J^{(1)}=\frac{z^4}{2}
\end{gather}
the evolution w.r.t.\ the time $t_0$ becomes
\begin{gather}
\partial_0 {h_{-1}}=-{\dfrac {\partial_x (h_{-1}) h_1}{{2h_{-1}}^{2}}}+
{\dfrac {\partial_x  h_1}{2h_{-1}}},
\nonumber\\
\partial_0 {h_{0}}=  -\dfrac{3}{4} \dfrac{(\partial_x h_{-1})^2 h_1}{{h_{-1}}^4}
+\dfrac{3}{4} \dfrac{(\partial_x h_{-1})(\partial_x
h_{1})}{{h_{-1}}^3} +\dfrac{1}{4} \dfrac{(\partial_x^2 h_{-1})
h_{1}}{{h_{-1}}^3} -\dfrac{1}{4} \dfrac{\partial_x^2
h_{1}}{{h_{-1}}^2},
\nonumber\\
\partial_0{h_1}={\dfrac {(\partial_x  h_{-1}){h_1}^{2}}{2{h_{-1}}^{3}}}
-{\dfrac {15}{8}}{\dfrac {h_1 (\partial_x
h_{-1})^{3}}{{h_{-1}}^{6}}} +{\dfrac {15}{8}}{\dfrac {(\partial_x
h_{-1} ) ^{2}\partial_x h_1 }{{h_{-1}}^{5}}}+\dfrac{5}{4}{\dfrac
{h_1(\partial_x^{2}h_{-1}) \partial_x  h_{-1}
 }{{h_{-1}}^{5}}}
\nonumber\\
\phantom{\partial_0{h_1}=} {}-\dfrac{3}{4}{\dfrac {(\partial_x
h_{-1})\partial_x^{2}  h_1 }{{h_{-1}}^{4}}}-{\dfrac {h_1\partial_x
h_1 }{{2h_{-1}}^{2}}} -{\dfrac {(\partial_x^{ 2 } h_{-1})
\partial_x  h_1 }{2{h_{-1}}^{4}}}-{\dfrac {h_1  \partial_x^3
h_{-1} }{8{h_{-1}}^{4}}}+{\dfrac {\partial_x^{3}h_1
}{8{h_{-1}}^{3}}}.\label{extlCH2}
\end{gather}
One can see that the f\/ield $h_0$ does not af\/fect the evolution
of the system. This is true also for the following times of the
reduction because the currents themselves do not depend on $h_0$.
Actually, by direct computation, we can show that the current
$J^{(0)}$ do not depend on $h_0$. By~(\ref{esco}) all the
successive currents do not depend on $h_0$.

The more compact way to write the system is
\begin{gather}
\partial_0 \alpha= D_{\alpha} \dfrac{\gamma}{\alpha},
\qquad
\partial_0 \beta= -D_{\alpha}^2 \dfrac{\gamma}{\alpha},
\qquad
\partial_0 \gamma= \alpha D_{\alpha}^3 \dfrac{\gamma}{\alpha},\label{lCH2}
\end{gather}
where $\alpha=h_{-1}$, $\gamma= h_{-1} h_{1}$, $\beta=h_{0}$ and
$D_{\alpha}=\partial_x \cdot (1/2\alpha)$.

The conserved densities for (\ref{lCH2}) are given by the
condition (\ref{constr}) written in terms of the generator $h$ of
the Fa\`a di Bruno polynomials (\ref{FdBz2}):
\begin{gather}
\label{quasiRiccati2}
 \frac{1}{2 \alpha^2} h^{(2)}
- \biggl(\frac{ \alpha_x}{2\alpha^3} +\frac{\beta}{\alpha^2}
\biggr) h^{(1)} + \biggl( \frac{\beta_x}{2\alpha^2}
+\frac{\gamma}{\alpha^2} - \frac{\beta  \alpha_x}{2\alpha^3} -
\frac{\beta^2 }{\alpha^2}\biggr)h^{(0)}=\frac{z^4}{2}.
\end{gather}
The local CH can be obtained from this equation system under the
constraint:
\begin{gather*}
\gamma= \frac{m_{xx}}{8 m}-\frac{5 m_{x}^2} {32 m^2}+\frac{1}{8},
\qquad \beta=-\partial_x \ln (m^{1/4}), \qquad \alpha=\sqrt{m}
\end{gather*}
which reduces (\ref{lCH2})  to (\ref{locCH}), i.e.:
\begin{gather*}
\partial_0 {m}=\frac{1}{4}\left( \partial_x-\partial_x^3 \right)\frac{1}{\sqrt{m}}.
\end{gather*}
The same reduction transforms  (\ref{quasiRiccati2}) into
(\ref{Riccati}). An important property of (\ref{lCH2}) is related
to its associated dispersionless system:
\begin{gather*}
\partial_0 \alpha= \dfrac{1}{2}\partial_x \dfrac{y}{\alpha^2},
\qquad
\partial_0 \beta=0,
\qquad
\partial_0 \gamma=0.
\end{gather*}
This trivial system is bi-Hamiltonian w.r.t. the Poisson tensors:
\begin{gather*}
P_1^d=\frac{1}{2}\left(\begin{matrix} \partial_x & 0& 0 \\0 & 0& 0
\\0 & 0& 0\end{matrix} \right), \qquad
P_0^d=\frac{1}{2}\left(\begin{matrix} 0 &0 & \partial_x \alpha\\
0 & 0& 0 \\\alpha \partial_x &0& \gamma \partial_x + \partial_x
\gamma \end{matrix} \right)
\end{gather*}
and the Hamiltonians $H_1= -\int({\gamma}/{\alpha})\,dx$ and
$H_0=\int ({\gamma^2}/{2\alpha^3})\,dx$. The metric associated to
$P_1^d$ is obviously degenerate, then this dispersionless system
is not of ``Dubrovin--Novikov type'' \cite{DN}. This is an
important dif\/ference between this reduction and the KdV equation
because it is not possible to include that system into the
Dubrovin's classif\/ication scheme.

\section{Conclusion}
In this paper we have studied the KP system related to the
Camassa--Holm hierarchy \cite{CLOP}. This system seems very
dif\/ferent from the standard KP case. The f\/irst dif\/ference is
in the fact that the evolution of the currents (\ref{esco}) is
not contained in the space $J_+$ which  plays the role of the
projection on the positive part of the Sato Grassmannian naturally
def\/ined by the currents.

The second dif\/ference is that the f\/irst reduction, obtained
f\/ixing the f\/irst current, is not an integrable system whose
dispersionless part is related to a non-degenerate metric. This
could be seen as a motivation why they do not enter in the
Dubrovin classif\/ication scheme \cite{DubZh}.

In the paper we have def\/ined, in a restricted case, a function $\psi$,
and we conjecture that its generalizations are the Baker--Akhiezer functions
for the KP-CH integrable reductions. We are studying also whether in our case
$\psi$ is an eigenfunction of a suitable Lax operator which can be
used to construct a zero curvature equation. The open question
that we will address is as follows: Are the local symmetries of CH
bilinear?

\subsection*{Acknowledgements}
The author thanks Marco Pedroni and Gregorio Falqui for very
useful discussions and hints, and two anonymous referees for
useful comments and bibliographical references.

\pdfbookmark[1]{References}{ref}
\LastPageEnding

\end{document}